\documentclass{Interspeech2024}
\usepackage{nvas3d}

\interspeechcameraready

\title{Novel-view Acoustic Synthesis From 3D Reconstructed Rooms}

\name{Byeongjoo}{Ahn}
\name{Karren}{Yang}
\name{Brian}{Hamilton}
\name{Jonathan}{Sheaffer}
\name{Anurag}{Ranjan}
\name{Miguel}{Sarabia}
\name{Oncel}{Tuzel}
\name{Jen-Hao Rick}{Chang}

\address{
  Apple}
\email{\{byeongjoo, karren\_yang, jenhao\_chang\}@apple.com}


\begin{document}

\maketitle

\begin{abstract}
We investigate the benefit of combining blind audio recordings with 3D scene information for novel-view acoustic synthesis.
Given audio recordings from 2--4 microphones and the 3D geometry and material of a scene containing multiple unknown sound sources, we estimate the sound anywhere in the scene.
We identify the main challenges of novel-view acoustic synthesis as sound source localization, separation, and dereverberation.
While naively training an end-to-end network fails to produce high-quality results, we show that incorporating room impulse responses (RIRs) derived from 3D reconstructed rooms enables the same network to jointly tackle these tasks.
Our method outperforms existing methods designed for the individual tasks, demonstrating its effectiveness at utilizing 3D visual information. 
In a simulated study on the Matterport3D-NVAS dataset, our model achieves near-perfect accuracy on source localization, a PSNR of \SI{26.44}{\dB} and a SDR of \SI{14.23}{\dB} for source separation and dereverberation, resulting in a PSNR of \SI{25.55}{\dB} and a SDR of \SI{14.20}{\dB} on novel-view acoustic synthesis.
We release our code and model on our project website at \url{https://github.com/apple/ml-nvas3d}. Please wear headphones when listening to the results.

\end{abstract}

\section{Introduction} \label{sec:intro}

Recent advancements in novel-view synthesis and 3D reconstruction~\cite{mildenhall2020nerf, nerfstudio, chang2023pointersect} have enabled users to explore scenes freely, viewing them from positions not captured during recordings.
However, a significant limitation of these approaches is the absence of sound, restricting the immersive experience.
In contrast to novel-view image synthesis, the non-stationary nature of sound and the low resolution of microphones make novel-view acoustic synthesis a challenging problem~\cite{chen2023novel}.

In this work, we investigate novel-view acoustic synthesis in 3D reconstructed and calibrated rooms.
We build upon recent developments in 3D reconstruction~\cite{mildenhall2020nerf} and acoustic calibration techniques~\cite{schissler2017acoustic, ratnarajah2022fast, ratnarajah2022mesh2ir} and assume the availability of high-quality room geometry and acoustic material information.
We also assume to have audio recordings from a limited microphone array (2--4 receivers) at known locations in the scene. 
However, we have no knowledge about the sound sources, including their number, locations, and content. 
Under this setting, our goal is to enable users moving freely in the scene to hear realistic spatial audio renderings of the unknown sound sources recorded by the microphones.

Novel-view acoustic synthesis remains challenging despite having 3D reconstructed rooms.
The main problem is the lack of knowledge of the sound sources. 
If we knew all information about the sound sources, including their locations and dry sound, given that we have a 3D reconstructed room (with room geometry and acoustic materials), we could synthesize audio at any new location using standard acoustic renderers~\cite{chen2020soundspaces, chen2022soundspaces, funkhouser1998beam, schissler2014high}. 
In other words, the key to novel-view acoustic synthesis is acoustic-scene reconstruction, \ie, estimating the positions (sound localization) and the content (sound separation and dereverberation) of the sound sources from blind audio recordings. 
However, the limited number and resolution of microphones and the mixture of different reverberant sound in same audios makes these problems difficult, 
particularly if the semantics are similar (\ie, two people speaking) or the sounds arrive from close directions. %
Existing methods relying on time-delay cues cannot pinpoint the locations of multiple sound sources in a complex 3D scene~\cite{michel2006history}.
Simply training a neural network also fails to generate high-quality results, as will be shown in \cref{sec:results}.

\begin{figure*}[t]
    \centering
    \includegraphics[width=\linewidth]{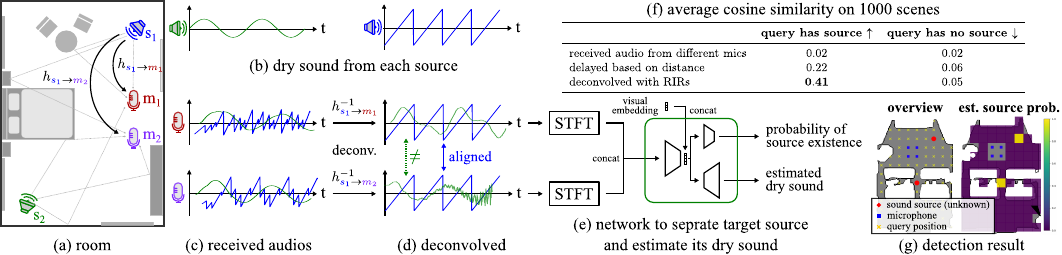}
    \caption{\textbf{Model overview and motivation.} Given a 3D reconstructed room (a) and audio recordings from microphones (c), 
    our method simultaneously performs sound source localization, separation, and dereverberation  to estimate the locations and dry sound of individual sound sources.
    (d,f) Our method utilizes a key observation that deconvolving audio recordings with the impulse response from a specific source location aligns sound emitted at that location across input recordings while keeping sound from other locations uncorrelated. Utilizing the aligned audios (while still mixed and reverberant) makes the problem easier for neural networks to perform said tasks.
    (e) We use a network to isolate target audio from the mixture of sounds and mitigate deconvolution artifacts.
    (g) Our source detection result on an example scene. Our network accurately identify where the sound sources are.
    }
    \label{fig:idea}
\end{figure*}

Our key observation is that the echoes caused by the multipath reflection of sound contain valuable information for acoustic scene reconstruction.
Specifically, as shown in \cref{fig:idea}, we propose to assist the network by deconvolving audio recordings from individual microphones (c) with RIRs from a specific location. 
This operation aligns the sound emitted from that location across microphone channels while keeping sound from other locations uncorrelated (d).  
Note that the deconvolved audios still contain mixed and reverberant sound from all sources --- only the sound component originated from the location specified by the RIR will be aligned and dereverberated in the individual recordings. 
This is a strong cue that enables neural network to predict whether an audio source exists at that location and perform sound separation and estimate the corresponding dry sound.
Iterating this approach over all candidate source locations enables us to reconstruct the acoustic scene with high spatial resolution (g) and render it from novel viewpoints using an acoustic renderer~\cite{chen2022soundspaces}.
Additionally, if we incorporate semantic visual cues (\emph{i.e.}, RGB images), %
we can further enhance source localization.  %
We thoroughly study the benefits of the proposed method, as well as the use of visual cues, in experiments on the Matterport3D-NVAS dataset.

\paragraph{Contributions}
Our method advances novel-view acoustic synthesis by leveraging RIRs for source localization, separation, and dereverberation. 
Our technique can reconstruct acoustic scenes with semantically indistinguishable sources, such as two guitars in the same room.
Additionally, as shown in our experiments, our model generalizes well to new scenes, as scene-specific information is encoded in the deconvolved audios.  %
We release our pretrained model and code in \href{https://github.com/apple/ml-nvas3d}{our project website}.

\section{Related Work} \label{sec:related}
Our approach relies on RIR estimation techniques that operate on 3D reconstructed rooms~\cite{schissler2017acoustic,ratnarajah2022fast,ratnarajah2022mesh2ir,tang2020scene,ratnarajah2023listen2scene,luo2022learning}. These techniques estimate or render RIRs by using both room geometry and estimated acoustic material properties.

There are extensive studies separately addressing the tasks of sound localization \cite{sarabia2023spatial}, separation \cite{rouard2023hybrid, mo2022closer,zhao2018sound,zhao2019sound,ephrat2018looking,michelsanti2021overview}, and dereverberation \cite{chen2023learning}. These generally do not take 3D scene information into account, which is useful for novel-view acoustic synthesis.
Although the active audio-visual separation method~\cite{majumder2022active} provides separation with 3D localization, it requires 3D embodied agents moving within the space. %

Conceptually, our method is related to beamforming techniques~\cite{nair2019audiovisual,gannot2017consolidated,michel2006history, teutsch2007modal,van1988beamforming, stoica2005spectral,van2002detection, thiergart2013geometry, mccormack2022object}, 
where the objective is to isolate the source at a query location, typically in non-reverberant scenarios (\eg, open spaces), relying on the directionality of sound.
Thiergart~\etal~\cite{thiergart2013geometry} and McCormack~\etal~\cite{mccormack2022object} utilize multiple microphone arrays or ambisonic microphones to triangulate the sound sources, assuming that any time-frequency tile in the spectrograms of the microphone arrays is from a single direct sound source.
In comparison, our method utilizes 3D scene geometry and performs dereverberation simultaneously. Our method supports complex acoustic scenes that contains sound sources with similar spectrograms (\eg, two guitars) playing concurrently, and it uses a small number (2-4) of omni-directional microphones.

Chen~\etal~\cite{chen2023sound} propose a self-supervised learning method to estimate camera rotation and sound source direction given an image and the binaural audio before camera rotation, and the image and a mono audio after camera rotation. Compared to our method, their proposed method focuses on estimating sound source direction with respect to the camera view and does not enable novel-view acoustic synthesis that allows free movement of the listeners.

Politis~\etal~\cite{politis2023widearea} propose an acoustic synthesis method that spatially interpolates the recorded audios from microphone arrays distributed across the scene. However, the location of the sound sources and listeners should be near the microphone arrays. In comparison, our method supports sound sources and listeners to be far away from the microphones.

Recently, Chen~\etal~\cite{chen2023novel} introduced ViGAS, a pioneering end-to-end approach for novel-view acoustic synthesis, using images to synthesize binaural audio. While their method supports novel listener locations away from input microphone and offers valuable insights, it does not address sound separation and is demonstrated only for a single source. It also does not utilize 3D scene geometry. In contrast, our method leverages 3D scene geometry for sound localization and separation, enabling it to handle multiple sources. %

\section{Method} \label{sec:method}
Our method decomposes novel-view acoustic synthesis into two subproblems: (i) acoustic scene reconstruction, which involves 3D source localization, separation, and dereverberation, and (ii) novel-view acoustic rendering. We begin by introducing each problem and then present our approach.

\subsection{Problem formulation}
\paragraph{Acoustic scene reconstruction}
Given recorded audio $y_m(t)$ from $M$ microphones and a 3D reconstructed room containing $S$ sound sources, we estimate the dry sound emitted by each source $\{x_s\}_{s=1}^S$, and their locations $\cP \, {=} \, \{\bp_s\}_{s=1}^S$. The recorded audio from microphone located at $\{\br_m\}_{m=1}^M$ is 
\begin{equation}
    y_m(t) = \sum_{s=1}^{S}\hsm(t) \ast x_s(t) + \psi_m(t),
\end{equation}
where $\hsm$ denotes the RIR from $\bp_s$ to $\br_m$, and $\psi_m(t)$ is the noise. Here we assume the RIRs are given from a 3D room reconstruction~\cite{schissler2017acoustic,tang2020scene,ratnarajah2022fast,ratnarajah2022mesh2ir,ratnarajah2023listen2scene,luo2022learning}.

\paragraph{Novel-view acoustic rendering}
Once the acoustic scene is known, novel-view acoustic rendering is straightforward as it can be achieved by simply convolving the dry sound results with corresponding RIRs (\eg, rendered by standard acoustic renderers~\cite{chen2022soundspaces}) for novel viewpoints. The audio $y(t)$ from novel microphone located at $\br$ is given by:
\begin{equation} \label{eq:render}
    y(t) = \sum_{s=1}^{S}h_{\bp_s \rightarrow \br}(t) \ast x_s(t).
\end{equation}

\subsection{Our approach}
\paragraph{Overview}
We reconstruct the acoustic scene by querying potential 3D source locations within a room. Our goal is to determine: (i) the existence of a source at a query location and (ii) if present, its associated dry sound. By iterating this process across potential source locations, we effectively localize, separate, and dereverberate all sources in the room. 

Specifically, for a set of candidate source locations, denoted as $\cQ \,{=}\, \{\bq_n\}_{n=1}^N$, which includes the actual source locations $\cP$ (\ie, $\cP \subset \cQ$), the network provides two outputs: (i) a detection estimate $\td$, indicating the presence of a source, $\mathbf{1}_{\mathcal{P}}(\bq_n)$, and (ii) an estimation $\tx_n(t)$ of the isolated dry sound $x_s(t)$ at the query point $\bq_n$ when a positive detection is made.
Then, novel-view acoustic synthesis is achieved by
\begin{equation}
    y(t) = \sum_{n=1}^{N}h_{\bq_n \rightarrow \br}(t) \ast \left(\tx_n(t) \ \mathbf{1}(\td>0.5)\right).
\end{equation}

\paragraph{Deconvolution and cleaning}
One challenge for processing multichannel audios from microphones at different locations is that the audios do not align across the channels. 
The delay and echo received by individual channels depend on source and microphone locations (\eg, see \cref{fig:idea}c) and can change dramatically across scenes.
Thus, a U-Net~\cite{gao2021visualvoice}, which performs well on single-channel source separation, fails with multiple channels when applied directly (see \cref{sec:results}).

Our key observation is that after deconvolving individual recorded audios with the RIR from a specific 3D location to a microphone, the sound emitted from the location (if any) would align across microphones. 
Specifically, given a query point $\bq_n$ and microphone $m$, we deconvolve the recorded audio $y_m(t)$ with the RIR $\hqm(t)$. 
In the frequency domain, the deconvolved audio can be represented as  
\begin{equation} \label{eq:deconvolution}
    Z_{nm}(w) = \sum_{s=1}^S \frac{\Hsm(w)}{\Hnm(w)} \ X_s(w) + \frac{\Psi_m(w)}{\Hnm(w)},
\end{equation}
where $X_s$, $\Hsm$, $\Hnm$, and $\Psi_m$ represent the Fourier transform of $x_s$, $\hsm$, $\hqm$, and $\psi_m$, respectively.

When $\bq_n$ corresponds to source $i$ (\ie, $\bq_n = \bp_i$), we have $Z_{nm}(w) = X_i(w) + \sum_{s\neq i} \frac{\Hsm(w)}{\Hnm(w)}X_s(w) + \frac{\Psi_m(w)}{\Hnm(w)}$, or $Z_{nm}(w) = X_i(w)$ + noise. 
Notice that $X_i$ is independent to $m$, which means that the deconvolved audios for individual microphones, $\{Z_{nm}\}_{m=1}^{M}$, consistently contain $X_i(w)$, the dry sound emitted by source $i$.
Sound from other sources are unaligned and become noise-like.
When $\bq_n$ contains no sound source, no such alignment would exist.
\cref{fig:idea}f shows the average cosine similarity between two microphone audios across $1000$ scenes with sources composed of speech and music. 
Deconvolution with RIRs significantly increases the similarity between two microphone channels when $\bq_n$ is a sound source and maintains low similarity otherwise.

Taking deconvolved audios as inputs, the neural network's job becomes separating the audios that are aligned across channels from the noise---a much easier job for a U-net than separating and dereverberating audios with arbitrary delay and echo.
In practice, we use Wiener deconvolution~\cite{wiener1964extrapolation} in our implementation to mitigate artifacts.

Our approach, which integrates deconvolution and cleaning, shares insights with image deblurring techniques that combine deconvolution with neural networks to reduce artifacts~\cite{xu2014deep}. Additionally, our method can be related to the basic delay-and-sum strategy in beamforming~\cite{teutsch2007modal,van1988beamforming}, but with key modifications: the traditional `delay' is substituted with RIR-based deconvolution, and the `sum' is replaced by a neural network to enable a more refined separation.

\paragraph{Utilizing visual information}
While our framework primarily relies on auditory cues across microphones, we can additionally utilize visual information. 
Specifically, we input the RGB environment map at the query location rendered from the 3D reconstructed room to the neural network.  %
When combined with deconvolved audio, it enhances the source detection and the final synthesis result, as shown in~\cref{sec:results}. 

\paragraph{Training}
For a query point $\bq$, our loss is composed of the Binary Cross-Entropy (BCE) for sound source detection and the Mean Squared Error (MSE) between the Short-Time Fourier Transforms (STFT) of the estimated dry sound $\tx$ and the ground truth dry sound $x_s$ when $\bq$ coincides a sound source:
\begin{equation}
\mathcal{L}(\bq_n) = \lambda\, \text{BCE}(\hat{d}, d) + d\, \Vert\text{STFT}(\hat{x}_n) - \text{STFT}(x_s)\Vert_2^2,
\end{equation}
where $\lambda$ is the detection weight, $d$ is the ground truth for source presence (1 if present), and $\td$ is the estimated probability of source existence. 
The MSE term is active solely for queries with a source. 
We use a U-Net architecture from VisualVoice~\cite{gao2021visualvoice} for source separation.
For source detection, we add a 3-layer CNN decoder, which takes the latent vector of the U-Net as input. 
For audio-visual experiments, we use a pretrained ResNet-18 to encode image features, which are concatenated with audio features at the latent space.

\section{Results} \label{sec:results}

\paragraph{Dataset} 
Following Chen~\etal~\cite{chen2023novel}, we create a simulated Matterport3D-NVAS dataset by using SoundSpaces~\cite{chen2020soundspaces,chen2022soundspaces} to render RIRs from Matterport3D scenes~\cite{chang2017matterport3d}, which are split into 51/11/11 rooms for train/validation/test sets. 
For sound sources, we incorporate speech recordings from the LibriSpeech dataset~\cite{panayotov2015librispeech} and audio from 12 MIDI instrument classes (bass, brass, chromatic percussion, drums, guitar, organ, piano, pipe, reed, strings, synth lead, synth pad) from the Slakh dataset~\cite{manilow2019cutting}, all sampled at 48kHz. %
We simulate RIRs between every pair of points on a \SI{1}{\m} resolution grid and randomly choose the placements of sound sources. We use all of the grid points as potential sound source locations (i.e., query points). 
For audio-visual experiments, we include a male or female mesh for LibriSpeech, and a guitar mesh for all instruments in Slakh.
For each scene, we randomly sample two source locations, paired with two random audios from LibriSpeech or Slakh.

\begin{table}[t] %
    \centering
    \caption{\textbf{Quantitative results.} %
    We evaluate our proposed method's capabilities of source localization (shown as detection), source separation (shown as reverberant sound), dereverberation (shown as dry sound) and novel-view acoustic synthesis (shown as NVAS). 
    } 
    \label{tab:quantitative}

    \setlength{\tabcolsep}{2pt}
    
    \begin{adjustbox}{max width=\linewidth}
    \scriptsize{
    \begin{tabular}{lccccccc}
        \toprule        
        \textbf{Method} & \textbf{Detection} & \multicolumn{2}{c }{\textbf{Dry sound}} & \multicolumn{2}{c}{\textbf{Reverberant sound}} & \multicolumn{2}{c}{\textbf{NVAS}} \\
        & AUROC$\,\uparrow$ & PSNR$\,\uparrow$ & SDR$\,\uparrow$ & PSNR$\,\uparrow$ & SDR$\,\uparrow$ & PSNR$\,\uparrow$ & SDR$\,\uparrow$ \\ 
        \cmidrule(r){1-1} \cmidrule(lr){2-2} \cmidrule(lr){3-4} \cmidrule(lr){5-6} \cmidrule(l){7-8} 
        {\scriptsize{\textbf{Comparisons (M=4)}}}\\
        Receiver audio & $0.507$  & $11.82$ & $-4.01$ & $13.13$ & $0.29$& $8.67$ & $3.39$ \\
        DSP w/o RIRs (delay-and-sum) &  $0.739$  & $13.06$ & $-0.97$ & $12.34$  & $0.06$ & $4.94$ & $3.59$  \\ 
        DSP w/ RIRs (deconv-and-sum) & $0.879$  & $15.79$ & $2.06$& $16.40$ & $2.69$ & $9.81$ & \cellcolor{orange!25}$6.09$ \\
        DL Baselines
        & \cellcolor{orange!25}$0.892$ & \cellcolor{orange!25}$16.54$ & \cellcolor{orange!25}$4.07$ & \cellcolor{orange!25}$23.87$ & \cellcolor{orange!25}$5.32$ & \cellcolor{orange!25}$22.77$ & $2.88$ \\
        Ours & \cellcolor{red!25}$0.996$ & \cellcolor{red!25}$26.28$ & \cellcolor{red!25}$14.14$ &  \cellcolor{red!25}$25.43$ & \cellcolor{red!25}$13.01$ & \cellcolor{red!25}$22.92$ & \cellcolor{red!25}$12.64$ \\ 
        \specialrule{.1em}{.05em}{.05em}\specialrule{.1em}{.05em}{.05em}
        {\scriptsize{\textbf{Ablation studies}}}\\
        Ours (M=3) & $0.985$ & $25.57$ & $13.34$ &  $25.01$ & $12.60$ & $21.42$ & $12.08$ \\ 
        Ours (M=2) & $0.987$  & $24.09$ & $11.68$ &  $24.02$ & $11.41$& $19.92$ & $10.28$ \\ 
        Ours (M=1) & $0.711$ & $16.29$ & $1.07$ &  $16.82$ & $1.91$ & $4.78$ & $0.17$ \\
        \cmidrule(r){1-1} \cmidrule(lr){2-2} \cmidrule(lr){3-4} \cmidrule(lr){5-6} \cmidrule(l){7-8} 
        Ours w/o deconv. (M=4) & $0.500$  & $10.95$ & $-7.92$ &  $10.85$ & $-7.75$& $12.53$ & $-\infty$ \\ 
        Ours w/o deconv. (M=4) + visual & $1.000$  & $11.01$ & $-8.34$ &  $10.91$ & $-8.14$& $11.38$ & $-2.36$ \\
        Ours (M=4) + visual &$1.000$  & $26.44$ & $14.23$ &  $25.64$ & $13.16$ &  $25.55$ & $14.20$ \\ 
        Ours (M=3) + visual & $1.000$ & $26.00$ & $13.80$ &  $25.38$ &$12.95$ & $24.93$ & $13.65$ \\
        Ours (M=2) + visual & $1.000$  & $24.23$ & $11.80$ & $24.12$ & $11.51$& $23.67$ & $12.60$ \\
        Ours (M=1) + visual& $1.000$ & $16.37$ & $1.28$ & $17.14$ & $2.50$ & $16.50$ & $6.04$ \\
        \bottomrule
    \end{tabular}
    }
    \end{adjustbox}
    \vspace{1.5em}
\end{table}

\paragraph{Tasks and metrics} We evaluate our method on novel-view acoustic synthesis, as well as on the intermediate tasks of source detection, separation and dereverberation (for acoustic scene reconstruction). 
For detection, we compute the area under the ROC curve (AUROC) based on source detection accuracy on the grid query points. 
For the other tasks, we use Peak Signal-to-Noise Ratio (PSNR) and Source-to-Distortion Ratio (SDR)~\cite{vincent2006performance} for evaluation.

\paragraph{Baselines and ablations}
Our method simultaneously tackles source localization, separation, and dereverberation. We compare with baselines designed for individual tasks.

\noindent \emph{DSP baselines:} Receiver signals are aligned using time-delay (DSP w/o RIRs) or deconvolution (DSP w/ RIRs). Thresholding based on cosine similarity is used for detection, and summation of signals is used for dry sound estimation. Reverberant sound and
novel-view sound are obtained by convolving dry sounds with
their corresponding RIRs.

\noindent  \emph{DL baselines:} We compare with recent learning-based methods designed for individual tasks---the network in~\cite{sarabia2023spatial} for detection, Demucs~\cite{rouard2023hybrid} for dry sound estimation, FUSS~\cite{wisdom2021s} for sound separation / reverberant sound estimation, and ViGAS~\cite{chen2023novel} for novel-view acoustic synthesis (NVAS). Demucs and FUSS use only a single microphone as input, and Demucs is evaluated only on speech enhancement. ViGAS does not support multiple sources in our data, thus we adapted their reported metrics to ours using their code.

\noindent  \emph{Ablations:} We evaluate the effect of deconvolution, visual information, and the number of microphones.

\paragraph{Acoustic scene reconstruction}
Table \ref{tab:quantitative} shows the quantitative comparison. 
Our method, which combines 3D scene information (via deconvolution with RIRs) with a learned network, outperforms all baselines on sound localization, separation, and dereverberation.

First, we demonstrate that it is important to deconvolve the input audios to the network. Without deconvolution, the same network fails on individual tasks (\emph{Ours}~v.~\emph{Ours w/o deconv}), and in general, there is a performance gap between models with and without deconvolution (\emph{i.e.,} compare \emph{Ours}~v.~\emph{DL Baselines} and \emph{DSP w/ RIRs}~v.~\emph{DSP w/o RIRs}). %
Without deconvolution, the task is very challenging since in addition to separate the sound from individual sources, the network needs to identify the time delays, which depends on the combination of microphone configurations, source locations, and the scene geometry.

Second, while deconvolution alone may be sufficient for dereverberating audios containing a single sound source, our experiments involve multiple overlapping sound sources, causing significant artifacts in the deconvolved receiver signals. 
Training a neural network to identify the aligned components within the deconvolved audios and map to the desired output is critical to performance (\emph{i.e.,} compare \emph{Ours}~v.~\emph{DSP w/ RIRs}). 
We include the source-separated dry sound results in the supplemental material.

Last, our method enables pinpointing multiple sound sources within complex 3D scenes, achieving near-perfect AUROC for detection, whereas existing methods are only capable of estimating directions and distances without consideration of scene geometry \cite{sarabia2023spatial}.

\begin{figure}[t] 
    \centering
    \includegraphics[width=1.0\linewidth]{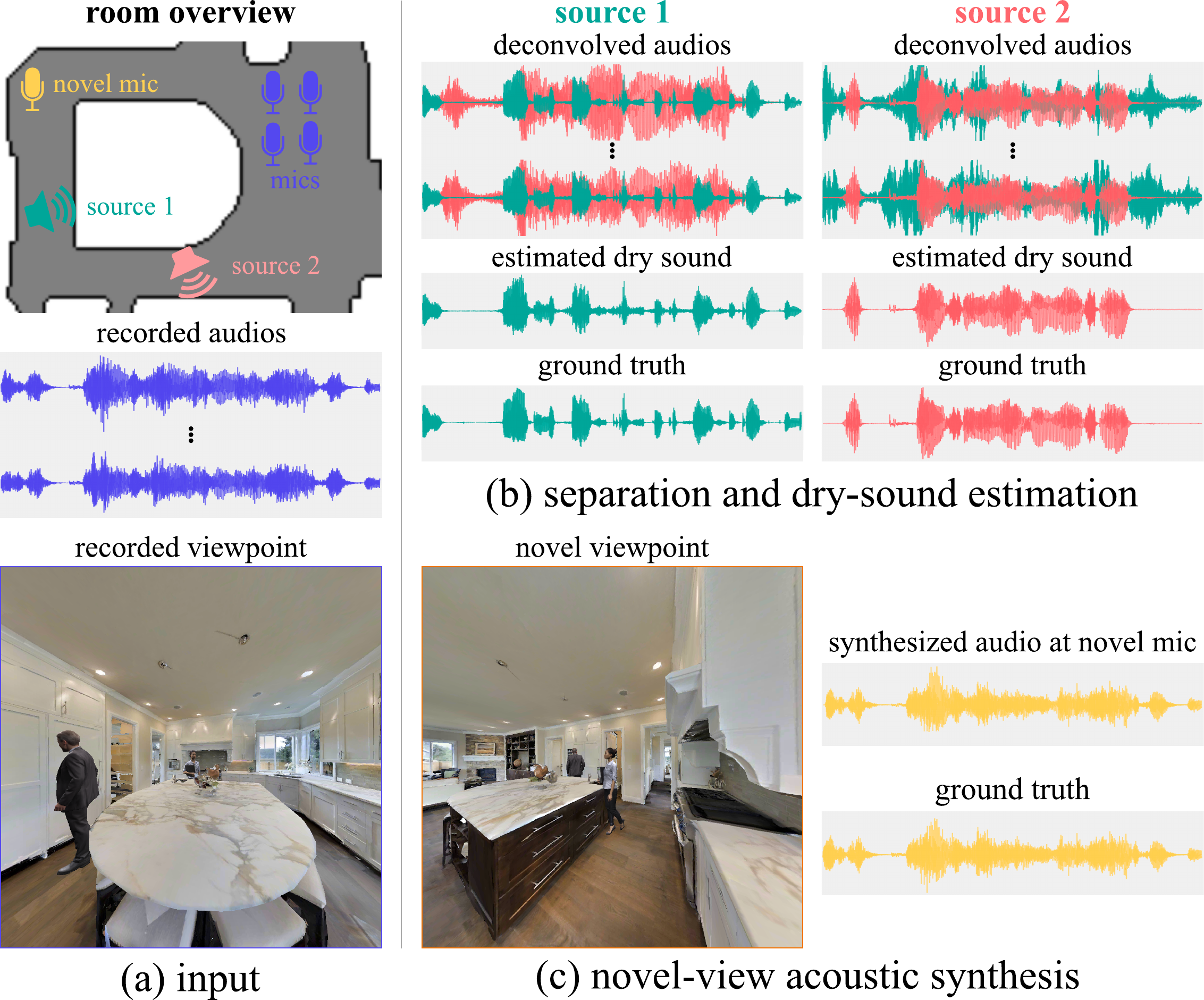}
    \caption{\textbf{Qualitative examples.} (a) Receiver audio is recorded and (b)  deconvolved with simulated RIRs. Leveraging the alignment from deconvolved audios, our method efficiently extracts dry sound, resulting in (c) synthesized audio from a novel viewpoint closely resembling true audio.}
    \label{fig:qualitative}
\end{figure}
\paragraph{Novel-view acoustic synthesis}
By localizing sound sources and estimating their dry sound, we effectively reconstruct the 3D acoustic scene. We subsequently resynthesize the audio anywhere in the scene using a standard renderer (SoundSpaces~\cite{chen2022soundspaces}). Our approach outperforms ViGAS \cite{chen2023novel}, which uses visual information and operates on single sound sources. Adding visual information further improves our results by boosting source detection accuracy and improving estimated dry sound (\emph{i.e.,} compare \emph{Ours}~v.~\emph{Ours+visual}).
Fig.~\ref{fig:qualitative} shows waveforms of the results for dry sound estimation and novel-view acoustic synthesis.
Our method effectively separates the dry sound from deconvolved audio and synthesizes novel-view audio consistent with ground truth. Please wear headphones and listen to the binaural NVAS results on our website.

\section{Conclusion} \label{sec:conclusion}

We investigated and introduced a method for novel-view acoustic synthesis which leverages advancements in 3D reconstruction and acoustic rendering. Our approach reconstructs an acoustic scene by jointly solving source localization, separation, and dereverberation, which allows us to realistically render immersive audio for users moving freely in a scene.
By deconvolving RIRs, our method utilizes 3D scene information and produces high quality source-separated dry sound that enables high quality novel-view acoustic synthesis compared to existing deep learning baselines.

\paragraph{Limitations}
We model sound sources as omni-directional point sources for simplicity. 
Our method relies on the availability of RIRs, and its performance can be influenced by the quality of the estimated RIRs.

\paragraph{Acknowledgement}
We thank Dirk Schroeder and David Romblom for insightful discussions and feedback, Changan Chen for the assistance with SoundSpaces.

\bibliographystyle{IEEEtran}
\bibliography{refs}

\end{document}